\documentclass[journal]{IEEEtran}
\usepackage{graphicx} \usepackage{bm} \usepackage{amsmath}
\usepackage{cite}
\usepackage{amssymb}

\usepackage{amsmath,amsthm,amssymb,mathrsfs}
\usepackage{bm}
\usepackage{subfigure}

\newcommand{\D}{{\rm d}}

\newcommand{\I}{{\rm i}}

\renewcommand{\Re}{{\rm Re }}
\renewcommand{\Im}{{\rm Im }}

\ifCLASSINFOpdf
\else
\fi
\begin{document}
\title{Penetration Depth of Transverse Spin Current in Ferromagnetic Metals}

\author{Tomohiro~Taniguchi${}^{1,2}$,%~\IEEEmembership{Member,~IEEE,}
        Satoshi~Yakata${}^{3}$,%~\IEEEmembership{Fellow,~OSA,}
        Hiroshi~Imamura${}^{1}$,%~\IEEEmembership{Fellow,~OSA,}
        and~Yasuo~Ando${}^{4}$%,~\IEEEmembership{Life~Fellow,~IEEE}% <-this % stops a space
        \\
        ${}^{1}$Nanotechnology Research Institute, 
        National Institute of Advanced Industrial Science and Technology,
        \\
        Tsukuba, Ibaraki 305-8568, Japan
        \\
        ${}^{2}$Institute of Applied Physics, University of Tsukuba, 
        Tsukuba, Ibaraki 305-8573, Japan
        \\
        ${}^{3}$Nanoelectronics Research Institute, 
        National Institute of Advanced Industrial Science and Technology,
        \\
        Tsukuba, Ibaraki 305-8568, Japan
        \\
        ${}^{4}$ Department of Applied Physics, Graduate School of Engineering, 
        Tohoku University, Sendai 980-8579, Japan
%\thanks{Tomohiro Taniguchi: tomohiro-taniguchi@aist.go.jp}% <-this % stops a space
\thanks{Manuscript received}}

\maketitle

\begin{abstract}
%\boldmath
The line width of the ferromagnetic resonance (FMR) spectrum of 
Cu/CoFeB/Cu/Co/Cu is studied. 
Analyzing the FMR spectrum by the theory of spin pumping,  
we determined the penetration depth of the transverse spin current in the Co layer. 
The obtained penetration depth of Co is 1.7 nm.
\end{abstract}


% \fi
\IEEEpeerreviewmaketitle

\section{Introduction}

% =========================================================================================== %

%\IEEEPARstart{T}{his} demo file is intended to serve as a ``starter file''

\IEEEPARstart{T}{here} is great interest in the field of
current-driven magnetization dynamics (CDMD) 
because of its potential applications to 
non-volatile magnetic random access memory and microwave devices. 
The concept of CDMD was first proposed 
by Slonczewski [1] and independently by Berger [2] in 1996. 
In the last decade, many experimental studies have shown the evidences of CDMD [3],[4]. 
% =========================================================================================== %

Theoretical studies of CDMD  have also been developped [5],[6]. 
The origin of the CDMD has been understood as 
the transfer of spin angular momentum 
of the conducting electrons to the magnetization of the ferromagnetic metal. 
One of the most important quantities in CDMD is 
the penetration depth of the transverse (perpendicular to the magnetization) 
spin current $\lambda_{\rm t}$, 
over which the transfer of spin angular momentum is achived. 
However, 
there is a controversial issue 
regarding the penetration depth of the transverse spin current. 
The ballistic theory of electron transport argues that 
$\lambda_{\rm t}$ is on the order of the lattice constant 
in conventional ferromagnets such as Fe, Co and Ni, and their alloys [7],[8]. 
On the other hand, 
the Boltzmann theory of electron transport argues that 
$\lambda_{\rm t}$ is on the order of a few nm [9],[10],[11]. 
However, only a few experimental measurements of the penetration depth
has been reported [12],[13].

% =========================================================================================== %

In our previous paper [14], 
we studied the line width of the ferromagnetic resonance (FMR) spectrum 
in a ferromagnetic(F)/nonmagnetic(N) metal five-layer system 
(N${}_{1}$/F${}_{1}$/N${}_{2}$/F${}_{2}$/N${}_{3}$), 
and showed that 
the line width of the F${}_{1}$ layer depends on the thickness of the F${}_{2}$ layer 
due to spin pumping [15],[16]. 
Analyzing the FMR spectrum, 
the penetration depth of the transverse spin current of NiFe, CoFe and CoFeB were obtained [14]. 
Our result seems to support the Boltzmann theory of electron transport. 
However, we cannot compare our results with the results of [9],[10],[11] directly 
since only the penetration depth of Co is studied in [9],[10],[11]. 
In this paper, we study the line width of FMR spectrum of 
Cu/CoFeB/Cu/Co/Cu five-layer system, 
and determine the penetration depth of Co. 
The obtained penetration depth of Co, 1.7 nm, 
has good agreement with the results of [9],[10],[11].

% =========================================================================================== %
% =========================================================================================== %

\begin{figure}
  \centerline{\includegraphics[width=0.8\columnwidth]{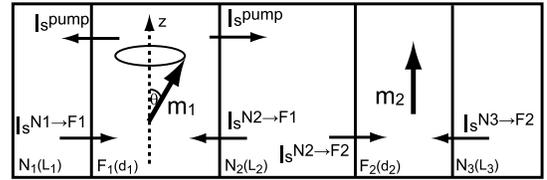}}
  \caption{
    Schematic illustration of a nonmagnetic \!\!/\!\!
    ferromagnetic metal five-layer system. 
    $L_{i}(i=1,2,3)$ is the thickness of the $i$-th nonmagnetic layer 
    and $d_{k}(k=1,2)$ is the thickness of the $k$-th ferromagnetic layer. 
    The magnetization $\mathbf{m}_{1}$ is in resonance 
    and precess around the $z$-axis with the angle $\theta$. 
    The magnetization $\mathbf{m}_{2}$ is fixed along the $z$-axis. 
    $\mathbf{I}_{s}^{\rm pump}$ and $\mathbf{I}_{s}^{{\rm N}\to{\rm F}}$ are 
    pumped spin current and backflow of spin current, respectively. 
  }
  \label{fig:fig1}
\end{figure}

% =========================================================================================== %
% =========================================================================================== %

\section{Theory}
% =========================================================================================== %
Spin pumping [15],[16] is, in some sense, 
the reverse process of CDMD, 
where the precession of the magnetization in the ferromagnetic layer 
generates spin current 
flowing into the adjacent layers. 
In a ferromagnetic/nonmagnetic metal multi-layer system 
the Gilbert damping constant of the ferromagnetic layer 
is enhanced due to spin pumping. 
Analyzing the dependence of the Gilbert damping 
on the thickness of the nonmagnetic layer 
the spin diffusion length, 
i.e., the penetration depth of spin current 
in the nonmagnetic layer is determined. 

% =========================================================================================== %

The penetration depth of the transverse spin current 
of a ferromagnetic metal, $\lambda_{\rm t}$, is 
also determined in a similar way. 
Let us consider N${}_{1}$/F${}_{1}$/N${}_{2}$/F${}_{2}$/N${}_{3}$ metal 
five-layer system shown in Fig. \ref{fig:fig1}, 
where $\mathbf{m}_{k}$ ($k\!\!=\!\!1,2$) is the unit vector 
along the magnetization of the $k$-th ferromagnetic layer. 
The magnetization of the F${}_{1}$ layer ($\mathbf{m}_{1}$) is in resonance 
with the oscillating magnetic field, 
and pumps spin current $\mathbf{I}_{s}^{\rm pump}$ 
flowing into the other layers. 
The precession axis of $\mathbf{m}_{1}$ is along 
the direction of the magnetization of the F${}_{2}$ layer ($\mathbf{m}_{2}$). 
Since the magnetization vector of $\mathbf{I}_{s}^{\rm pump}$ is 
perpendicular to $\mathbf{m}_{1}$ [14] 
and the precession angle $\theta$ is very small (about 1 deg), 
the dominant component of the magnetization vector of spin current 
flowing into the F${}_{2}$ layer is perpendicular to $\mathbf{m}_{2}$, 
i.e., 
the dominant component of the spin current flowing into the F${}_{2}$ layer 
is the transverse spin current. 
Thus, 
analyzing the dependence of the FMR spectrum of the F${}_{1}$ layer 
on the thickness of the F${}_{2}$ layer, 
the penetration depth of the transverse spin current of the F${}_{2}$
layer can be determined. 
However, 
the conventional theory of spin pumping 
assumes that the penetration depth of the transverse spin current is zero. 
Thus, we need to extend the theory of spin pumping 
by taking into account the finite penetration depth [14]. 

% =========================================================================================== %
The spin current pumped from the F${}_{1}$ layer is given by [15]
\begin{equation}
  \mathbf{I}_{s}^{\rm pump}
  \!=\!
  \frac{\hbar}{4\pi}
  \left(
    g_{{\rm r}({\rm F}_{1})}^{\uparrow\downarrow}
    \mathbf{m}_{1}\!\times\!\frac{\D\mathbf{m}_{1}}{\D t}
    +
    g_{{\rm i}({\rm F}_{1})}^{\uparrow\downarrow}\frac{\D\mathbf{m}_{1}}{\D t}
  \right)\ ,
  \label{eq:pump}
\end{equation}
where $\hbar$ is the Dirac constant 
and $g_{\rm r(i)}^{\uparrow\downarrow}$ is the real (imaginary) part of the mixing conductance. 
The pumped spin current creates spin accumulation in the other layers. 
The spin accumulation is given by 
\begin{equation}
  \bm{\mu}
  =
  \int_{\varepsilon_{\rm F}}\!\! \D \varepsilon 
  {\rm Tr}[\hat{\bm{\sigma}}\hat{f}]\ ,
  \label{eq:accumulation}
\end{equation}
where $\hat{\bm{\sigma}}$ is the Pauli matrix 
and $\hat{f}$ is the non-equilibrium distribution matrix in spin space. 
In general, 
the distribution matrix of a ferromagnetic layer, $\hat{f}_{\rm F}$, is given by 
$\hat{f}_{\rm F}
  \!=\!
  f_{0}\hat{1}
  +
  f_{z}\mathbf{m}\cdot\hat{\bm{\sigma}}
  +
  f_{x}\mathbf{t}_{1}\cdot\hat{\bm{\sigma}}
  +
  f_{y}\mathbf{t}_{2}\cdot\hat{\bm{\sigma}}$,
where 
$\hat{1}$ is the $2\times 2$ unit matrix, 
$f_{0}\!=\!(f^{\uparrow}\!+\!f^{\downarrow})\!/2$ is 
the non-equilibrium charge distribution and 
$f_{z}\!=\!(f^{\uparrow}\!-\!f^{\downarrow})/2$ is 
the difference in non-equilibrium distribution 
between spin-up ($f^{\uparrow})$ and spin-down ($f^{\downarrow}$) electrons. 
($\mathbf{t}_{1},\mathbf{t}_{2},\mathbf{m}$) is a set of orthogonal
unit vectors in spin space where $\mathbf{m}$ is the unit vector
parallel to the magnetization vector. 
$f_{x}$ and $f_{y}$ are 
the non-equilibrium distribution of the transverse spin components. 
Spin accumulation in a nonmagnetic layer is defined in a similar way. 

% =========================================================================================== %
The spin accumulation induces 
a backflow of spin current. 
The backflow of spin current 
flowing from the N${}_{i}$ layer to the F${}_{k}$ layer 
is expressed in terms of 
the spin accumulation as 
\begin{equation}
\begin{split}
  \mathbf{I}_{s}^{{\rm N}_{i}\to{\rm F}_{k}}
  =&
  \frac{1}{4\pi}
  \left[
    \frac{2g_{({\rm F}_{k})}^{\uparrow\uparrow}g_{({\rm F}_{k})}^{\downarrow\downarrow}}
      {g_{({\rm F}_{k})}^{\uparrow\uparrow}+g_{({\rm F}_{k})}^{\downarrow\downarrow}}
    \{\mathbf{m}_{k}\cdot(\bm{\mu}_{{\rm N}_{i}}-\bm{\mu}_{{\rm F}_{k}})\}\mathbf{m}_{k}
  \right.
\\
  &
  + g_{{\rm r}({\rm F}_{k})}^{\uparrow\downarrow}\mathbf{m}_{k}\times(\bm{\mu}_{{\rm N}_{i}}\times\mathbf{m}_{k})
  + g_{{\rm i}({\rm F}_{k})}^{\uparrow\downarrow}\bm{\mu}_{{\rm N}_{i}}\times\mathbf{m}_{k}
\\
  &
  \left.
  - t_{{\rm r}({\rm F}_{k})}^{\uparrow\downarrow}\mathbf{m}_{k}\times(\bm{\mu}_{{\rm F}_{k}}\times\mathbf{m}_{k})
  - t_{{\rm i}({\rm F}_{k})}^{\uparrow\downarrow}\bm{\mu}_{{\rm F}_{k}}\times\mathbf{m}_{k}
  \right]\ ,
  \label{eq:backflow}
\end{split}
\end{equation}
where 
$\bm{\mu}_{{\rm N}_{i}}$ and $\bm{\mu}_{{\rm F}_{k}}$ are 
the spin accumulation of the N${}_{i}$ and the F${}_{k}$ layer, respectively. 
$g^{\uparrow\uparrow(\downarrow\downarrow)}$ is the spin-up (spin-down) conductance 
and $t_{\rm r(i)}^{\uparrow\downarrow}$ is the real (imaginary) part of 
the transmisson mixing conductance 
defined at the F/N interface. 
In the conventional theory of spin pumping, 
the penetration depth of the transverse spin current 
is assumed to be zero, 
and the last two terms in (\ref{eq:backflow}) is neglected [8]. 

% =========================================================================================== %
The spin current given by (\ref{eq:pump}) and (\ref{eq:backflow}) satisfies
the boundary conditions of the continuity of the spin current. 
In general, 
the current operator in spin space is given by [9]
\begin{equation}
  \hat{j}
  =
  \frac{1}{e}\hat{C}\frac{\partial V}{\partial x}
  -
  \hat{D}\frac{\partial\hat{n}}{\partial x}\ ,
  \label{eq:current_operator}
\end{equation}
where $e(>0)$ is the absolute value of electron charge and 
$V$ is the applied voltage. 
Since we are interested in the FMR line width, we assume $V=0$.
$\hat{C}$, $\hat{D}$ and $\hat{n}$ are 
the $2\times 2$ matrices representing 
the conductivity, the diffusion constant 
and the density of the non-equilibrium electron, respectively. 
The conductivity and the diffusion constant are expressed as 
$\hat{C}\!=\!C_{0}(\hat{1}+\beta\mathbf{m}\cdot\hat{\bm{\sigma}})$ and 
$\hat{D}\!=\!D_{0}(\hat{1}+\beta^{'}\mathbf{m}\cdot\hat{\bm{\sigma}})$, 
where 
$C_{0}\!=\!(\sigma^{\uparrow}\!+\!\sigma^{\downarrow})/2$, 
$D_{0}\!=\!(D^{\uparrow}\!+\!D^{\downarrow})/2$, 
$\beta\!=\!(\sigma^{\uparrow}\!-\!\sigma^{\downarrow})/(\sigma^{\uparrow}\!+\!\sigma^{\downarrow})$ 
and 
$\beta^{'}\!=\!(D^{\uparrow}\!-\!D^{\downarrow})/(D^{\uparrow}\!+\!D^{\downarrow})$.
$\sigma^{\uparrow(\downarrow)}$ and $D^{\uparrow(\downarrow)}$ are 
the conductivity and the diffusion constant of 
spin-up (spin-down) electrons, respectively. 
$\beta$ and $\beta^{'}$ 
are the polarization of the spin dependent conductivity and diffusion constant, respectively. 
The conductivity and the diffusion constant satisfy 
the Einstein relation $\hat{C}\!=\!e^{2}\hat{N}\hat{D}$, 
where $\hat{N}$ is the density of states. 
For simplicity, 
we assume that $\beta\!\!=\!\!\beta^{'}$ in this paper. 
The distribution $\hat{f}$ and the density $\hat{n}$ are related with each other via
%\begin{equation}
%  {\rm Tr}[\hat{n}]
%  =
%  \int_{\varepsilon_{\rm F}}\D\varepsilon
%  {\rm Re}\left[
%          {\rm Tr}\left[
%                    \hat{N}\hat{f}
%                  \right]
%          \right]\ ,
%  \label{eq:density_distribution1}
%\end{equation}
\begin{equation}
  {\rm Tr}[\hat{\bm{\sigma}}\hat{n}]
  =
  \int_{\varepsilon_{\rm F}}\!\!\D\varepsilon
  {\rm Re}\left[
          {\rm Tr}\left[
                    \hat{\bm{\sigma}}\hat{N}\hat{f}
                  \right]
          \right]\ .
  \label{eq:density_distribution2}
\end{equation}
%The charge current is given by 
%$I_{c}=-eS{\rm Re}[{\rm Tr}[\hat{j}]]$, 
%where $S$ is the cross section area. 
%Similarly, 
The spin current is given by 
$\mathbf{I}_{s}\!=\!(\hbar S/2){\rm Re}[{\rm Tr}[\hat{\bm{\sigma}}\hat{j}]]$, 
where $S$ is the cross section area. 
Using (\ref{eq:accumulation}), (\ref{eq:current_operator}) and (\ref{eq:density_distribution2}), 
the spin current $\mathbf{I}_{s}$ is expressed 
in terms of spin accumulation $\bm{\mu}$. 
The spin current in a nonmagnetic metal is expressed in a similar way, 
but $\beta\!=\!\beta^{'}\!=\!0$. 

The diffusion equation of the spin accumulation is obtained by 
the continuity of the charge and spin current. 
In a nonmagnetic metal, 
the spin accumulation $\bm{\mu}_{\rm N}$ obeys the diffusion equation given by [17]
\begin{equation}
  \frac{\partial^{2}}{\partial x^{2}}\bm{\mu}_{\rm N}
  =
  \frac{1}{\lambda_{\rm sd(N)}^{2}}\bm{\mu}_{\rm N}\ ,
  \label{eq:diff_eq_N}
\end{equation}
where $\lambda_{\rm sd(N)}$ is the spin diffusion length of the nonmagnetic metal. 
The spin accumulation can be expressed as a linear combination of 
$\exp(\pm x/\lambda_{\rm sd(N)})$. 
The longitudinal spin accumulation in a ferromagnetic metal, 
$\bm{\mu}_{\rm F}^{\rm L}=(\mathbf{m}\cdot\bm{\mu}_{\rm F})\mathbf{m}$, 
also obeys the diffusion equation, 
and is expressed as a linear combination of $\exp(\pm x/\lambda_{\rm sd(F_{L})})$, 
where $\lambda_{\rm sd(F_{L})}$ is the longitudinal spin diffusion length. 

% =========================================================================================== %
We assume that the transverse spin accumulation in a ferromagnetic metal, 
$\bm{\mu}_{\rm F}^{\rm T}=\mathbf{m}\times(\bm{\mu}_{\rm F}\times\mathbf{m})$, 
obeys the following equation [9]:
\begin{equation}
  \frac{\partial^{2}}{\partial x^{2}}\bm{\mu}_{\rm F}^{\rm T}
  =
  \frac{1}{\lambda_{J}^{2}}\bm{\mu}_{\rm F}^{\rm T}\times\mathbf{m}
  +
  \frac{1}{\lambda_{\rm sd(F_{T})}^{2}}\bm{\mu}_{\rm F}^{\rm T}\ ,
  \label{eq:diff_eq_FT}
\end{equation}
where $\lambda_{J}=\sqrt{(D^{\uparrow}+D^{\downarrow})\hbar/(2J)}$ is 
the spin coherence length [8] 
and $\lambda_{\rm sd(F_{\rm T})}=\lambda_{\rm sd(F_{L})}/\sqrt{1-\beta^{2}}$ is 
the transverse spin diffusion length. 
$J$ represents the strength of the exchange field. 
The transverse spin accumulation is expressed as a linear combination of 
$\exp(\pm x/l_{+})$ and $\exp(\pm x/l_{-})$, 
where $1/l_{\pm}=\sqrt{(1/\lambda_{\rm sd(F_{T})}^{2})\mp(\I/\lambda_{J}^{2})}$. 
Therefore, we define the penetration depth of the transverse spin current as
\begin{equation}
  \frac{1}{\lambda_{\rm t}}
  =
  {\rm Re}\left[\frac{1}{l_{+}}\right]\ .
\end{equation}
References [10],[11] show that the order of $\lambda_{J}$ is a few nm 
for NiFe and Co. 
The exchange interaction, which determines $\lambda_{J}$, 
does not give any contribution to $\lambda_{\rm sd(F_{L})}$, 
i.e., there's no relation between $\lambda_{J}$ and $\lambda_{\rm sd(F_{L})}$. 
If the order of $\lambda_{\rm sd(F_{L})}$ is a few nm, for example NiFe, 
$\lambda_{\rm t}\!\!\simeq\!\!\lambda_{\rm sd(F_{L})}$. 
On the other hand, 
for Co $\lambda_{\rm sd(F_{L})}\!\!\gg\!\!\lambda_{J}$, and therefore 
$\lambda_{\rm t}\!\!\ll\!\!\lambda_{\rm sd(F_{L})}$.

% =========================================================================================== %
The spin current at the F${}_{2}$/N${}_{2}$(N${}_{3}$) interface is given by 
$\mathbf{I}_{s}^{{\rm N}_{2}\to{\rm F}_{2}}$ ($-\mathbf{I}_{s}^{{\rm N}_{3}\to{\rm F}_{2}}$). 
Soving the diffusion equations of spin accumulations 
of the N${}_{3}$ and F${}_{2}$ layers, (\ref{eq:diff_eq_N}) and (\ref{eq:diff_eq_FT}), 
with these boundary conditions, 
the backflow at the N${}_{2}$/F${}_{2}$ interface can be expressed as [13]
\begin{equation}
\begin{split}
   &
   \mathbf{I}_{s}^{{\rm N}_{2}\to{\rm F}_{2}}
   =
   \frac{1}{4\pi}
   \left[
     \tilde{g}_{({\rm F}_{2})}^{*}
     (\mathbf{m}_{2}\cdot\bm{\mu}_{\rm N_{2}})\mathbf{m}_{2}
   \right.
\\
  & \left. +
    \tilde{g}_{{\rm r}({\rm F}_{2})}^{\uparrow\downarrow}
    \mathbf{m}_{2}\times(\bm{\mu}_{\rm N_{2}}\times\mathbf{m}_{2})
    +
    \tilde{g}_{{\rm i}({\rm F}_{2})}^{\uparrow\downarrow}
    \bm{\mu}_{\rm N_{2}}\times\mathbf{m}_{2}
  \right]\ ,
\end{split}
\end{equation}
where the conductance $\tilde{g}_{({\rm F}_{2})}^{*}$ depends on 
the ratio $d_{2}/\lambda_{\rm sd(F_{L})}$, 
where $d_{2}$ is the thickness of the F${}_{2}$ layer. 
Similarly, 
the renormalized mixing conductances, $\tilde{g}_{{\rm r,i}({\rm F}_{2})}^{\uparrow\downarrow}$, 
depend on the ratio $d_{2}/l_{+({\rm F}_{2})}$. 
If the thickness of the N${}_{3}$ layer is thin enough 
compared to its spin diffusion length, 
$\tilde{g}_{({\rm F}_{2})}^{*}$ is equal to $g^{*}$ given in [14],
and $\tilde{g}_{{\rm r,i}({\rm F}_{2})}^{\uparrow\downarrow}$ are given by 
\begin{equation}
  \begin{pmatrix}
    \tilde{g}_{{\rm r}({\rm F}_{2})}^{\uparrow\downarrow} \\
    \tilde{g}_{{\rm i}({\rm F}_{2})}^{\uparrow\downarrow}
  \end{pmatrix}
  =
  \frac{1}{\Delta}
  \begin{pmatrix}
    K_{1} & K_{2} \\
    -K_{2} & K_{1}
  \end{pmatrix}
  \begin{pmatrix}
    g_{{\rm r}({\rm F}_{2})}^{\uparrow\downarrow} \\
    g_{{\rm i}({\rm F}_{2})}^{\uparrow\downarrow}
  \end{pmatrix}\ ,
  \label{eq:renormalized_mixing}
\end{equation}
where $\Delta=K_{1}^{\ 2}+K_{2}^{\ 2}$. 
$K_{1}$ and $K_{2}$ are given by 
\begin{equation}
  K_{1}
  =
  1
  +
  t_{\rm r}^{\uparrow\downarrow}\eta_{\rm r}
  +
  t_{\rm i}^{\uparrow\downarrow}\eta_{\rm i}\ ,
\end{equation}
\begin{equation}
  K_{2}
  =
  -t_{\rm r}^{\uparrow\downarrow}\eta_{\rm i}
  +
  t_{\rm i}^{\uparrow\downarrow}\eta_{\rm r}\ ,
\end{equation}
where 
$\eta_{\rm r(i)}=\Re(\Im)\eta$ 
and $\eta=\{g_{\rm t}\tanh(d_{2}/l_{+})\}^{-1}$, 
where $g_{\rm t}=hS/(2e^{2}\rho_{\rm F_{2}}l_{+})$, 
and $\rho_{\rm F_{2}}$ is the resistivity of the F${}_{2}$ layer. 
The mixing conductance of the F${}_{1}$ layer 
in (\ref{eq:pump}) and (\ref{eq:backflow}) is 
also replaced by the renormalized mixing conductance. 

% =========================================================================================== %
The spin pumping modifies the Landau-Lifshitz-Gilbert (LLG) equation 
of the magnetization of the F${}_{1}$ layer as 
\begin{equation}
  \frac{\D\mathbf{m}_{1}}{\D t}
  =
  -\gamma\mathbf{m}\times\mathbf{B}_{\rm eff}
  +
  \bm{\tau}
  +
  \alpha_{0}\mathbf{m}_{1}\times\frac{\D\mathbf{m}_{1}}{\D t}\ ,
  \label{eq:llg}
\end{equation}
where $\mathbf{B}_{\rm eff}$ is the effective magnetic field, 
$\gamma$ is the gyromagnetic ratio and 
$\alpha_{0}$ is the intrinsic Gilbert damping constant. 
$\bm{\tau}$ is the additional torque due to the spin pumping given by 
\begin{equation}
  \bm{\tau}
  =
  \frac{\gamma}{MSd_{1}}
  \mathbf{m}_{1}\times
  \{(\mathbf{I}_{s}^{\rm pump}-\mathbf{I}_{s}^{{\rm N}_{2}\to{\rm F}_{1}})\times\mathbf{m}_{1}\}\ ,
\end{equation}
where $M$ is the saturated magnetization of the F${}_{1}$ layer 
and $d_{1}$ is the thickness of the F${}_{1}$ layer. 
We assume that the spin relaxation in the N${}_{2}$ layer is so weak 
that the spin current in the N${}_{2}$ layer is conserved, 
i.e., $\mathbf{I}_{s}^{\rm pump}-\mathbf{I}_{s}^{{\rm N}_{2}\to{\rm F}_{1}}=\mathbf{I}_{s}^{{\rm N}_{2}\to{\rm F}_{2}}$. 
Then the dynamics of the magnetization of the F${}_{1}$ layer is affected by the F${}_{2}$ layer. 
We notice that 
the effects of the N${}_{1}$ and N${}_{3}$ layer are quite small 
because, as mentioned below, 
the thickness of these layers are thin enough 
compared to its spin diffusion length in our experiments. 
The LLG equation (\ref{eq:llg}) is rewritten as [14],[15],[18]
\begin{equation}
  \frac{\D\mathbf{m}_{1}}{\D t}
  =
  -\gamma_{\rm eff}\mathbf{m}_{1}\times\mathbf{B}_{\rm eff}
  +
  \frac{\gamma_{\rm eff}}{\gamma}
  (\alpha_{0}+\alpha^{'})
  \mathbf{m}_{1}\times\frac{\D\mathbf{m}_{1}}{\D t}\ ,
\end{equation}
where ($\gamma_{\rm eff}/\gamma$) and $\alpha^{'}$ 
is the enhancement of the gyromagnetic ratio and the Gilbert damping 
due to the spin pumping, respectively.
Assuming that $g_{\rm r}^{\uparrow\downarrow}\gg g_{\rm i}^{\uparrow\downarrow}$, 
in the limit of $\theta\to 0$, 
$\alpha^{'}$ is reduced as 
\begin{equation}
  \alpha^{'}
  \simeq
  \frac{\gamma\hbar}{4\pi Md_{1}S}
  \frac{\tilde{g}_{{\rm r}({\rm F}_{1})}^{\uparrow\downarrow}\tilde{g}_{{\rm r}({\rm F}_{2})}^{\uparrow\downarrow}}
    {\tilde{g}_{{\rm r}({\rm F}_{1})}^{\uparrow\downarrow}+\tilde{g}_{{\rm r}({\rm F}_{2})}^{\uparrow\downarrow}}\ ,
  \label{eq:alpha_prime}
\end{equation}
and $(\gamma_{\rm eff}/\gamma)\simeq 1$. 
We should note that 
if we neglect the penetration depth of the transverse spin current 
in the ferromagnetic layer 
the mixing conductances are not renormalized, 
and that the enhancement of the Gilbert damping constant, 
$\alpha^{'}$, does not depend on the thickness of the F${}_{2}$ layer. 
This is because the dominant component of the pumped spin current 
is perpendicular to the magnetization of the F${}_{2}$ layer.

% =========================================================================================== %
% =========================================================================================== %

\begin{figure}
  \centerline{\includegraphics[width=0.8\columnwidth]{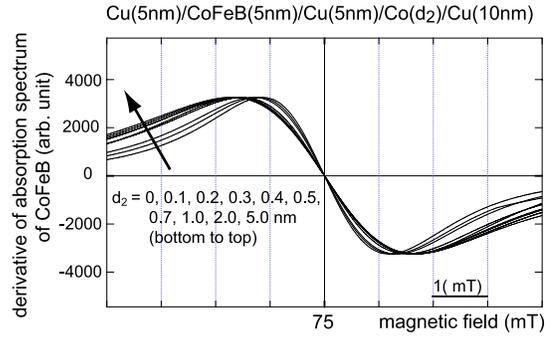}}
  \caption{
  The dependence of derivative of the FMR spectrum of CoFeB on the thickness of Co layer, $d_{2}$. 
  The center of the horizontal axis, 75 mT, is the resonance magnetic field of CoFeB. 
  }
  \label{fig:fig2}
\end{figure}

% =========================================================================================== %
% =========================================================================================== %

\section{Experiment}

We performed FMR experiments 
on Cu(5nm)/CoFeB(5nm) /Cu(5nm)/Co($d_{2}$)/Cu(10nm) five-layer system 
shown in Fig. \ref{fig:fig1} [16], 
where CoFeB layer corresponds to the F${}_{1}$ layer 
and Co layer corresponds to the F${}_{2}$ layer. 
Figure \ref{fig:fig2} shows 
the dependence of the derivative of the FMR spectrum of CoFeB 
on the thickness of Co, $d_{2}$. 
The width of the peak to peak in Fig. \ref{fig:fig2}, 
namely the linewidth of the FMR spectrum $\Delta B$, 
is a linear function of the Gilbert damping constant [19]:
\begin{equation}
  \Delta B
  =
  \Delta B_{0}
  +
  \frac{4\pi f}{\sqrt{3}\gamma}(\alpha_{0}+\alpha^{'})\ ,
  \label{eq:line_width}
\end{equation}
where $f$ is the frequency of the oscillating magnetic field. 
The line width of CoFeB depends on 
the thickness of Co 
through $\alpha^{'}$, as shown in Fig. \ref{fig:fig2}. 
Thus, we can determine the penetration depth of the transverse spin current of Co 
by the line width of CoFeB. 
The enhancement of the gyromagnetic ratio 
does not give any contributions to the line width.

% =========================================================================================== %

The sample was deposited on Corning 1737 glass substrates 
using an rf magnetron sputtering system 
in an ultrahigh vacuum below $4\times 10^{-6}$ Pa and cut to 5 nm${}^{2}$. 
The Ar pressure during deposition was 0.077 Pa. 
The FMR measurement was carried out 
using an X-band microwave source ($f\!=\!9.4$[GHz]) 
at room temperature. 
The microwave power, 
modulation frequency, 
and modulation field are 
1 mW, 10 kHz, and 0.1 mT, respectively. 
The precession angle of the magnetization of the F${}_{1}$ layer was estimated to be 1 deg. 
The resistivity of CoFeB and Co are 
1252 $\Omega\cdot$nm and 210 $\Omega\cdot$nm [20], 
respectively. 
The magnetization ($4\pi M$) 
and the gyromagnetic ratio 
of CoFeB are 1.66 T and 1.846$\times 10^{11}$ Hz/T, respectively.

% =========================================================================================== %

A Cu layer typically shows an enhanced (111) orientation 
and the Co layer on it also shows an induced (111) texture.
Thus, 
the Co layer is considered to be (111) texture.

% =========================================================================================== %
% =========================================================================================== %

\begin{figure}
  \centerline{\includegraphics[width=0.8\columnwidth]{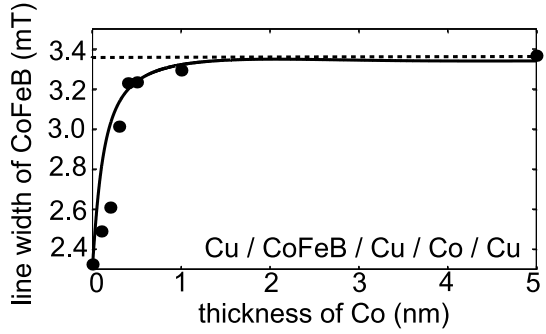}}
  \caption{
    The dependence of the FMR spectrum, $\Delta B$, of CoFeB layer 
    on the thickness of Co layer, $d_{2}$. 
    The filled circles represent experimental data 
    and solid line is fit to the experimental data 
    according to the theory with the finite penetration depth 
    of the transverse spin current in Co, $\lambda_{\rm t}$. 
    The dotted line represents the case of $\lambda_{\rm t}=0$. 
  }
  \label{fig:fig3}
\end{figure}

% =========================================================================================== %
% =========================================================================================== %

% =========================================================================================== %
In Fig. \ref{fig:fig3}
the measured line width of the FMR, $\Delta B$, of CoFeB layer 
is plotted with full circles 
against the thickness of Co layer, $d_{2}$. 
The solid line is a fit to the experimental data 
according to the theory with the finite penetration depth of
the transverse spin current $\lambda_{\rm t}$. 
The dotted line is the calculated line width assuming $\lambda_{\rm t}=0$. 
If the Co film is not continuous but consists of a Co islands, 
the thickness of the Co island is somewhat thicker than the nominal thickness. 
However, the effect of the Co islands is not so significant 
because the important quantity in our analysis is the mean thickness 
which is almost same as the nominal thickness.

% =========================================================================================== %
The best fitting parameters are as follows. 
The real part of the mixing conductances per unit area, 
$g_{\rm r}^{\uparrow\downarrow}/S$, 
of CoFeB and Co are 128 nm${}^{-2}$ and 20 nm${}^{-2}$, respectively. 
Although these values are determined by fitting, 
they have good agreement with 
the \textit{ab\ initio} calculations [8]. 
For simplicity, 
we assume that $t_{\rm r}^{\uparrow\downarrow}\!=\!t_{\rm i}^{\uparrow\downarrow}$, 
where the values of $t_{\rm r,i}^{\uparrow\downarrow}/S$ of 
CoFeB and Co are 0.8 nm${}^{-2}$ and 6.0 nm${}^{-2}$,respectively. 
The spin diffusion length of 
CoFeB and Co layer are 
12 nm and 38 nm, respectively [20],[21]. 
The polarization of the conductance $\beta$ are 
0.56 for CoFeB and 0.31 for Co [20],[21]. 
We take 
$g_{\rm i}^{\uparrow\downarrow}\!/\! S\!=\! 1.0$ nm${}^{-2}$ and 
$2g^{\uparrow\uparrow}g^{\downarrow\downarrow}\!/\!(g^{\uparrow\uparrow}\!+\!g^{\downarrow\downarrow}\!)S=20$ nm${}^{-2}$ 
both CoFeB and Co [15]; 
these are not important parameters for fitting. 
The spin diffusion length and resistivity of Cu are taken to be 500 nm and 21 $\Omega\cdot$nm [22]. 

% =========================================================================================== %
The obtained value of the penetration depth 
of Co is $\lambda_{\rm t}=1.7$ nm. 
References [9],[10],[11] estimate $\lambda_{\rm t}=\sqrt{2}\lambda_{J}$, 
and predict that 
$\lambda_{J}$ of Co with (111) texture is 1.1 nm. 
Thus, we have good agreement with [9],[10],[11].

\section{Conclusion}

In conclusion, 
we study the line width of the FMR spectrum of 
Cu/CoFeB/Cu/Co/Cu five-layer system. 
The line width of the CoFeB layer depends on the thickness of the Co layer 
due to spin pumping. 
We extend the conventional theory of spin pumping 
by taking into account the finite penetration depth of 
the transverse spin current of the Co layer, 
and analize the experimental data. 
The obtained penetration depth of the Co layer is 1.7 nm, 
which has good agreement with the Boltzmann theory of electron transport.

% or


%\appendices

%\section{}

\section*{Acknowledgment}

This work was supported by NEDO. 
One of the authors (T.T.) is supported by 
Research Fellowship of Japan Society for
the Promotion of Science 
for Young Scientist.

\ifCLASSOPTIONcaptionsoff
  \newpage
\fi

\end{document}